# Rare Earth Doping and Effective Band-Convergence in SnTe for Improved Thermoelectric Performance

Somnath Acharya[1], Dibyendu Dey[2], Tulika Maitra,[3] Ajay Soni[1*] and A. Taraphder[1,2*]

[1]School of Basic Sciences, Indian Institute of Technology Mandi, HP 175005 India

[2]Department of Physics and Centre for Theoretical Studies, Indian Institute of Technology Kharagpur 721302, India.

[3]Department of Physics, Indian Institute of Technology Roorkee, UK, 247667 India.

Corresponding Author: ajay@iitmandi.ac.in and arghya@phy.iitkgp.ernet.in

Thermoelectric performance of SnTe has been found to enhance with isovalent doping of alkaline and transition metal elements where most of these elements have solubility of less than 13%. We propose a strategy of doping rare earth element Yb to enhance the thermoelectric performance of SnTe. With heavy atomic mass and strong spin-orbit coupling, even the mild doping of Yb (~ 5%) is enough to create a degeneracy via band-convergence which enhances the density of states near Fermi level and improve overall performance. Our transport data and first-principles calculations corroborate that nearly 5% Yb is an efficient dopant to achieve thermoelectric response which is equivalent to 9% of Mn doping. The results are useful for understanding the environment-friendly thermoelectric SnTe.

**Introduction**

Recently, SnTe has attracted much attention as a non-toxic and environment friendly thermoelectric (TE) material over lead-based materials.[1] The hole doping from inherent Sn vacancies brings the charge carriers in the metallic region, thus the Seebeck coefficient is poor. Consequently, doping in SnTe has been used as an effective strategy to control charge carriers for TE application.[1-5] Technologically, TE materials can directly and reversibly convert waste heat to electrical energy for power generation and refrigeration with acoustically silent and emission free techniques.[6] The performance of any TE material depends on the dimensionless figure of merit, $ZT = (S^2\sigma/\kappa)T$, where $S$ is Seebeck coefficient, $\sigma$ is electrical conductivity, $\kappa$ is total thermal conductivity which has contributions from both electronic ($\kappa_e$) and lattice ($\kappa_l$) parts, and $T$ is average absolute temperature, respectively.[6-8] The interdependence of $S$, $\sigma$ and $\kappa_e$ with carrier concentration, '$n$' is a major hurdle for improving overall performance as $ZT$ and power factor, $S^2\sigma$ usually peak in a narrow range of '$n$'.[8] The prime challenge for TE application of SnTe is the intrinsic Sn vacancies which lead to high '$n$' and hence extremely low $S$ and high $\kappa$.[4-6] Further, having an analogous electronic band structure to PbTe, the SnTe also has contributions from both the valence bands of light (VB$_1$) and heavy (VB$_2$) holes. However, the energy difference (E$_b$) between the VB$_1$ and VB$_2$ valence bands in SnTe is too large compared to the same in PbTe, resulting in lesser contribution of heavy hole to $S$.[9-12] Due to these unfavourable factors, SnTe has a substantially limited TE performance. In the last few years, several elements (transition and alkaline metals) have been employed for doping in SnTe, not only to improve the electrical properties by (i) valence band-convergence [2, 13-17], (ii) introduction of resonant states near Fermi level ($E_F$)[18-20] and (iii) optimization of carrier concentration, [13, 21, 22] but also to obtain poor thermal responses via (i) formation of secondary phase [13, 14, 23], (ii) point defect engineering [4, 22, 24] and (iii) appearance of soft phonon modes.[5] In most of these cases, the better TE response is obtained with a higher level ($\geq 9\%$) of doping at the Sn sites.[2, 13-15, 25] Using a simple valence band model, much higher values of $S$ at room temperature have been shown for doped SnTe samples in comparison for Pisarenko plots.[18] Recently, we have also demonstrated that the effect of interaction between magnetic dopants and charge carriers to be an unique concept for enhancement of the carrier effective mass (m*) and TE performance of SnTe.[3] These results indicate that TE performance of SnTe can be improved with doping and it can emerge as an alternative to PbTe.

Rare earth elements have heavy atomic mass and it is expected that their doping could substantially enhance the density of states (DOS) near $E_F$ to improve $S$.[26-28] In addition, the local interaction from $f$-electrons may induce additional carrier scattering.[26] For instance, doping with rare earth Yb has improved the electronic transport properties by enhancing the DOS near $E_F$ in PbTe[28] and other binary compounds ($YbAl_2$, $YbAl_3$).[29] Thus doping with rare earth element in SnTe might enhance its TE performance. Owing to a good solubility level of Yb in SnTe, we have tested effect of Yb doping on thermoelectric properties of SnTe.[30, 31] Therefore, we report experimental observation of a three-time increment in $S^2\sigma$ for SnTe with Yb doping (only 5%), which provides efficient degeneracy of the valence bands $VB_1$ and $VB_2$ as supported by our first principle calculations. Further, the first principles calculation shows that due to heavy atomic mass (~173.04 g mol$^{-1}$) and strong spin-orbit coupling, Yb doping (~ 5%) can converge the two valence bands more effectively than Mn at the same doping level. We have achieved a better $S^2\sigma$ for $Sn_{0.95}Yb_{0.05}Te$ in comparison to earlier reports on Mn doped SnTe.[3, 5] Thus doping with rare earth elements is an efficient alternative to improve TE performance of SnTe.

**Experimental details**

High purity Sn powder (99%), Yb chunks (99.9%) and Te shot (99.99%) were used for preparing polycrystalline bulk sample of $Sn_{1-x}Yb_xTe$ (x = 0, 0.03 and 0.05) by solid state melting reaction. Stoichiometric amount of elements were placed in vacuum sealed (~ 10$^{-5}$ mbar) quartz tubes. To a brief note, the sealed tubes were heated slowly to 723 K in 12 hrs and then heated up to 1273 K in 5 hrs followed by soaking there for 10 hrs. The obtained ingots were crushed, ground into powders for cold pelletization followed by annealing at 773 K for 72 hrs. We refer to our earlier work[4] for phase purity characterizations of these compounds. With the clean and single phase of XRD data the samples are shown to be of good quality. More details of preparation and characterization details were described in our earlier work.[3-5] High temperature $S$ and $\sigma$ measurements were performed on a bar shape (~ 3 mm x 2 mm x 10 mm) samples in a homemade setup, where $S$ was measured by differential method with $\Delta T = 10$ K and $\sigma$ in conventional four probe geometry under high vacuum. Quantum design physical property measurements system was used for low temperature transport and Hall Effect measurements.

**Theoretical Methods**

First-principles calculations based on the density functional theory (DFT) [32, 33] were performed by using the plane-wave basis as implemented in the Vienna ab-initio simulation package (VASP).[34, 35] We used projector augmented wave (PAW)[36, 37] potentials in our calculations and the wave functions were expanded in the plane-wave basis with a kinetic energy cut off 500 eV. Moreover, we also used generalized gradient approximation (GGA) with the Perdew-Burke-Ernzerhof (PBE)[38] parametrization for the exchange-correlation energy functional. Total energies were converged to less than $10^{-5}$ eV and reciprocal space integration was carried out with a Γ-centered k-mesh of 7 x 7 x 7. Spin-orbit coupling (SOC) was incorporated in our calculations for the parent and doped compounds within GGA+SO approximation.

**Results and discussion**

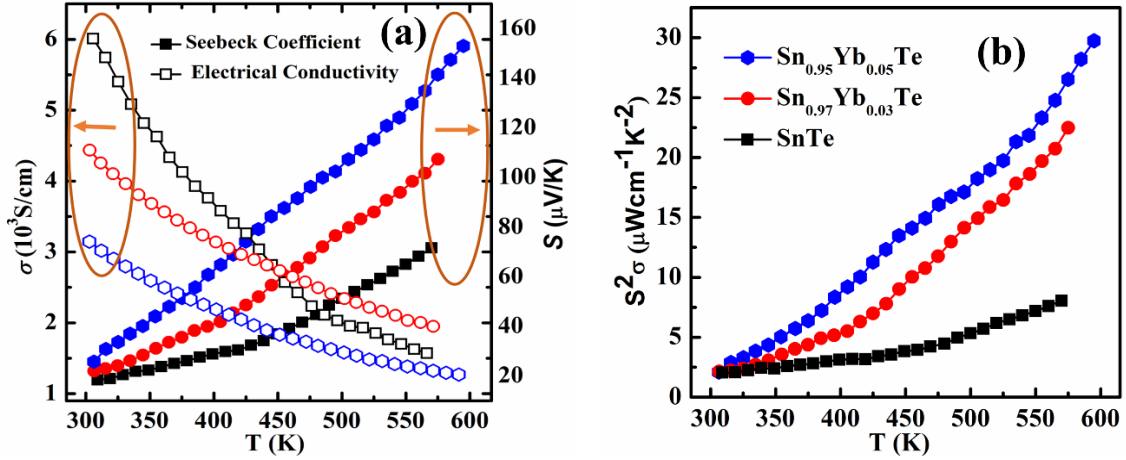

**FIG. 1.** Temperature dependence of (a) electrical conductivity, σ (left axes, open symbols) and Seebeck coefficient, $S$ (right axes) and (b) power factor, $S^2\sigma$ of $Sn_{1-x}Yb_xTe$ in the temperature range 305-600 K. Color codes are described in (b).

To understand the effect of Yb doping on the transport properties of SnTe, we have measured the temperature dependent σ, $S$, and $S^2\sigma$ in the temperature range 305-600 K. Shown in Fig. 1(a), σ decreases with the increase of temperature for all the samples, presenting the expected metallic behavior. However, Yb-doping decreases σ from ~ 6006 S/cm for SnTe to ~ 3145 S/cm for $Sn_{0.95}Yb_{0.05}Te$, at 305 K and this reduction of σ in doped samples is attributed to lower μ values. The estimated '$n$' and mobility, '$μ$' at room temperature are tabulated in Table 1, for series of $Sn_{1-x}Yb_xTe$. For Yb doped samples, the scattering is dominated by the disorders due to doping and the effect of temperature is subdominant and weaker while in the pure compound, the temperature dominates scattering and therefore, the electrical conductivity of the pure compound comes down below the doped sample. The cross over seems to be common at high temperatures, which is also observed in other reports on doped SnTe.[2, 14, 16] Further, the positive $S$ value enhances with temperature for all Yb-doped samples (Fig. 1 (a)). For example, at 575 K, the $S$ ~71.6 μV/K (for SnTe) enhances to ~139.65 μV/K (for $Sn_{0.95}Yb_{0.05}Te$), which is nearly double the value. The enhancement of $S$ is predictable from the well-known valence band engineering effect.[2, 13, 15] Convergence of the two valence bands would increase the density of states effective mass $m_d^*$ (decrease the $μ$) of bands near the $E_F$ and improve $S$ values, significantly.[9] Here, the band degeneracy for high $S$ is achieved with only 5% of rare earth Yb doping, which reveals that Yb is an effective dopant in SnTe. Though Yb doping decreases σ in the entire temperature range, the enhancement in $S$ is compensating the detrimental effect on the overall transport properties of $Sn_{1-x}Yb_xTe$. To confirm the hypothesis of improved transport properties, the temperature dependent $S^2\sigma$ has been calculated in the temperature range 305-595

K (shown in Fig. 1 (b)). The $S^2\sigma$ in Yb doped SnTe shows significant enhancement in the entire temperature range. For instance, listed in table 1, the $S^2\sigma$ value of $Sn_{0.95}Yb_{0.05}Te$ sample around room temperature is 2.09 $\mu Wcm^{-1}K^{-2}$ and at 575 K, it reaches 26.21 $\mu Wcm^{-1}K^{-2}$, which is more than three-time of pristine SnTe (~ 8.04 $\mu Wcm^{-1}K^{-2}$ at 575 K). The $S^2\sigma$ for 5 % doped Yb reaches 29.75 $\mu Wcm^{-1}K^{-2}$ at 595 K, which is a notable enhancement for single-doped SnTe at the same temperature.

To further understand the role of Yb in transport properties of doped SnTe, we have shown in Fig. 2 the Yb composition dependent $\sigma$, $S$, $\kappa$ and $S^2\sigma$ in the temperature range 100-575 K. As discussed, in the high temperature region, $\sigma$ decreases with temperature for $Sn_{1-x}Yb_xTe$, suggesting metallic behaviour for all the samples (shown in Fig. 2 (a)). For pure SnTe, $\sigma$ decreases from ~12378 S/cm to ~1570 S/cm in the temperature range 100-575 K, while 5% Yb doped SnTe shows nearly five times reduction of $\sigma$ (~1325 S/cm from ~6477 S/cm). Contrary to $\sigma$, the $S$ value increases with Yb content at all temperatures, as shown in Fig. 2 (b). For example, at room temperature $S$ reaches ~ 30 $\mu V/K$ (for $Sn_{0.95}Yb_{0.05}Te$) from ~19 $\mu V/K$ (for SnTe), which is a 56% enhancement over pure SnTe. The positive $S$ indicates that the majority charge carriers are holes, supporting the hall measurement results (Table 1). Here, the enhancement of $S$ with Yb doping can be understood with added heavy hole mass contribution by the decrease of energy gap between light and heavy hole valence bands (discussed later). Further, shown in Fig. 2 (c), the $\kappa$ value of $Sn_{0.95}Yb_{0.05}Te$ (~2.94 $Wm^{-1}K^{-1}$) is significantly low in comparison to SnTe (~ 4.46 $Wm^{-1}K^{-1}$) at room temperature. According to earlier report,[4] the reduction of $\kappa$ values with Yb doping is mainly due to enhancement of phonon scattering which decreases the effective phonon mean free path via interfacial grain boundary and point defect scattering. As shown in Fig. 2 (d), combining the enhanced $S$ with $\sigma$, all the doped samples exhibit higher $S^2\sigma$ than the pristine SnTe in the entire temperature range. Moreover, with the combination of electrical and thermal transport data, we have estimated the $ZT$ ~ 0.04 at 350 K for 5 % Yb doped SnTe (double of the pure sample), which is comparable to ~ 8% Mg doped ($ZT$ ~ 0.025) and ~ 16% Sb doped ($ZT$ ~ 0.05) SnTe.[2, 23]

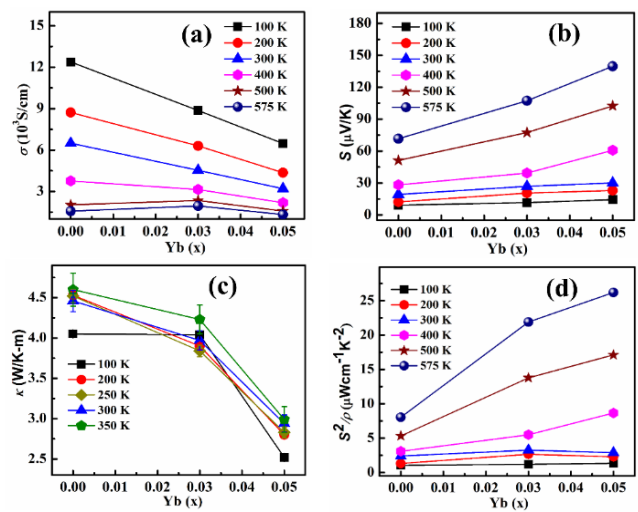

**Fig. 2**. Yb composition dependence of (a) Conductivity, $\sigma$, (b) Seebeck coefficient, $S$ (c) thermal conductivity, $\kappa$ and (d) power factor, $S^2\sigma$ at different temperature.

**Table 1**. Carrier concentration, *n*, Hall mobility, *μ*, and power factor, $S^2\sigma$ for $Sn_{1-x}Yb_xTe$.

| Compounds | $n$ ($10^{20}cm^{-3}$) at 300 K | $\mu$ ($cm^2 V^{-1}s^{-1}$) at 300 K | $S^2\sigma$ ($\mu Wcm^{-1}K^{-2}$) at 300 K | $S^2\sigma$ ($\mu Wcm^{-1}K^{-2}$) at 575 K |
|---|---|---|---|---|
| SnTe | 1.99 | 203.94 | 2.41 | 8.04 |
| $Sn_{0.97}Yb_{0.03}Te$ | 1.55 | 182.45 | 3.26 | 21.9 |
| $Sn_{0.95}Yb_{0.05}Te$ | 3.74 | 53.56 | 2.89 | 26.2 |

To understand the effect of Yb, the valence band-convergence has also been demonstrated by first-principles calculations based on the DFT. In the parent compound, light hole valence band (VB$_1$) maximum appears at the L point of the Brillouin Zone (BZ) (Fig. 3a) and the heavy hole valence band (VB$_2$) is located slightly below (~ 0.26 eV) VB$_1$. The band-convergence took place when the $E_b$ is of the order of the thermal excitation ($k_BT$). The band-convergence increases the m$_d$* efficiently and hence, a significant improvement in the TE performance is observed. Here, E$_b$ also decreases because of doping with other elements including alkaline or transition metal.

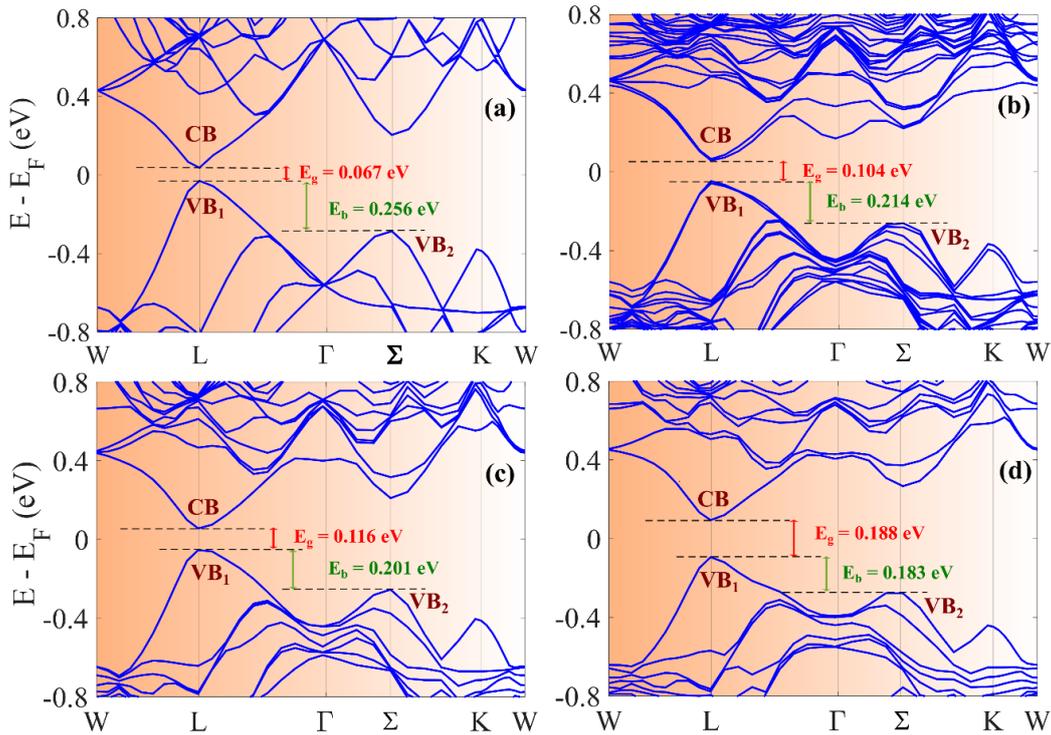

**FIG. 3.** (Color online) Calculated band structure for (a) SnTe, (b) $Sn_{0.97}Mn_{0.03}Te$, (c) $Sn_{0.97}Yb_{0.03}Te$, and (d) $Sn_{0.93}Yb_{0.07}Te$ within GGA+SO calculations. The E$_g$ observed at the L-point and VB$_2$ maximum appears at the Σ point.

Our ab-initio band-structure calculations (Fig.3) shows that the band-convergence effect is more prominent with rare earth (Yb) doping than transition metal doping (Mn as representative element). Within our GGA+SO calculations the direct band-gap ($E_g$) at the L-point is 0.067 eV (Fig.3 (a)) for the parent SnTe, which is consistent with earlier results,[39] while $E_b$ at the L and $\Sigma$ points is 0.256 eV. By substituting 1/27 Sn with Yb, we get $Sn_{0.97}Yb_{0.03}Te$ and perform our calculations. The $VB_1$ maximum in Fig.3(c) shifts downwards, and the conduction band minimum shifts upwards, which results in an enhancement of $E_g$ to ~ 0.116 eV. On the other hand, the $VB_2$ maximum at the $\Sigma$ point is much more rigid against the substitution, and there is hardly any shift of the peak position. Therefore, $E_b$ is reduced to ~ 0.201 eV, primarily due to the band-gap enhancement. A higher content of Yb (2/27) at the Sn sites can enhance $E_g$ significantly to ~ 0.188 eV and reduces $E_b$ to ~ 0.183 eV (Fig. 3(d)). Now, if SnTe is doped with transition metal Mn ($Sn_{0.97}Mn_{0.03}Te$), the band-convergence also takes place. It is evident from Fig. 3(b) that $E_g$ enhances to ~ 0.104 eV and consequently $E_b$ reduces to ~ 0.214 eV. Nevertheless, $E_g$ ($E_b$) is less (high) for $Sn_{0.97}Mn_{0.03}Te$ than for $Sn_{0.97}Yb_{0.03}Te$ (Fig. 3c).

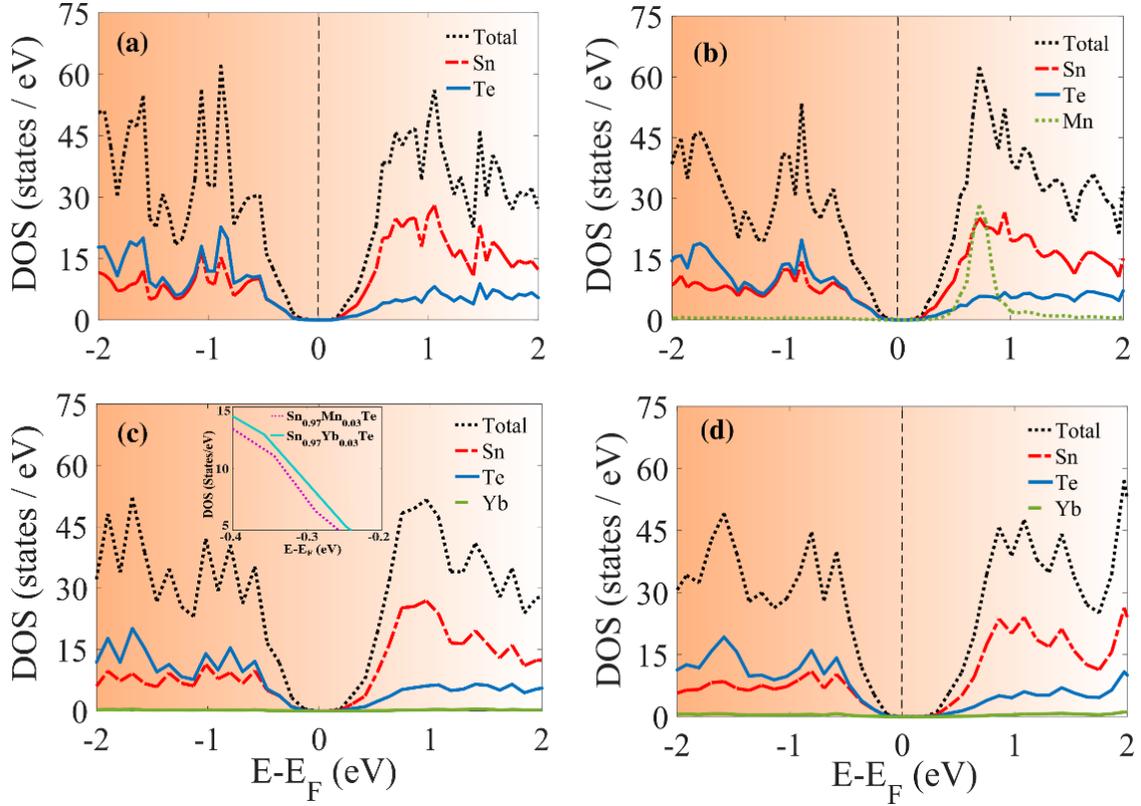

**FIG. 4.** (Color online) Electronic density of state of (a) SnTe, (b) $Sn_{0.97}Mn_{0.03}Te$, (c) $Sn_{0.97}Yb_{0.03}Te$, and (d) $Sn_{0.93}Yb_{0.07}Te$ within GGA+SO calculations. The enhancement of DOS with Yb doping (over Mn doping) at ~ 0.3 eV below $E_F$ in the inset of (c).

When the $E_F$ (for the experimental carrier concentration) closer to the $VB_2$ maximum falls on larger DOS, $S$ has been enhanced significantly. If we look at Fig.4 (a-d), sharp rise in the DOS can be observed around ~ 0.3 eV below the $E_F$ in all the samples, where the band-convergence takes place. The DOS is significantly larger in the Yb-doped compounds (Inset in Fig 4c). A detailed comparison between the DOS of the parent and doped compounds implies that Yb-doping achieves higher enhancement in the DOS compared to Mn-doping in SnTe for the same concentration. It can be observed from Fig. 4(d) that further increment of Yb-content enhances the DOS as the heavy-hole valence band becomes flat. The reason behind better band-convergence due to Yb-doping can be understood in terms of SOC. The SOC is stronger in heavier elements as the corresponding energy $E_{SO}$ is proportional to $Z^4$, where Z is the atomic number. In this case, Yb is a rare-earth element (Z=70) whereas Mn is a transition metal element (Z=25) and the effect of SOC is more prominent in the Yb-doped compounds, which leads to enhancement in the $E_g$ and pushes the $VB_1$ nearer to $VB_2$ which is shown in Fig.5 (a) and (b). The schematic diagrams, shown in Fig. 5 (c) demonstrates the band convergence of the two bands. We have also compared the $S^2\sigma$ of pure SnTe and $Sn_{0.94}Mg_{0.09}Te$,[2] $Sn_{0.98}Bi_{0.02}Te+3\%HgTe$,[14] $Sn_{0.94}Ca_{0.09}Te$,[15] $Sn_{0.91}Mn_{0.09}Te$[16] and $Sn_{0.93}Mn_{0.1}Te$[3] in Fig. 5 (d), where an equivalent enhanced performance is observed for $Sn_{0.95}Yb_{0.05}Te$.

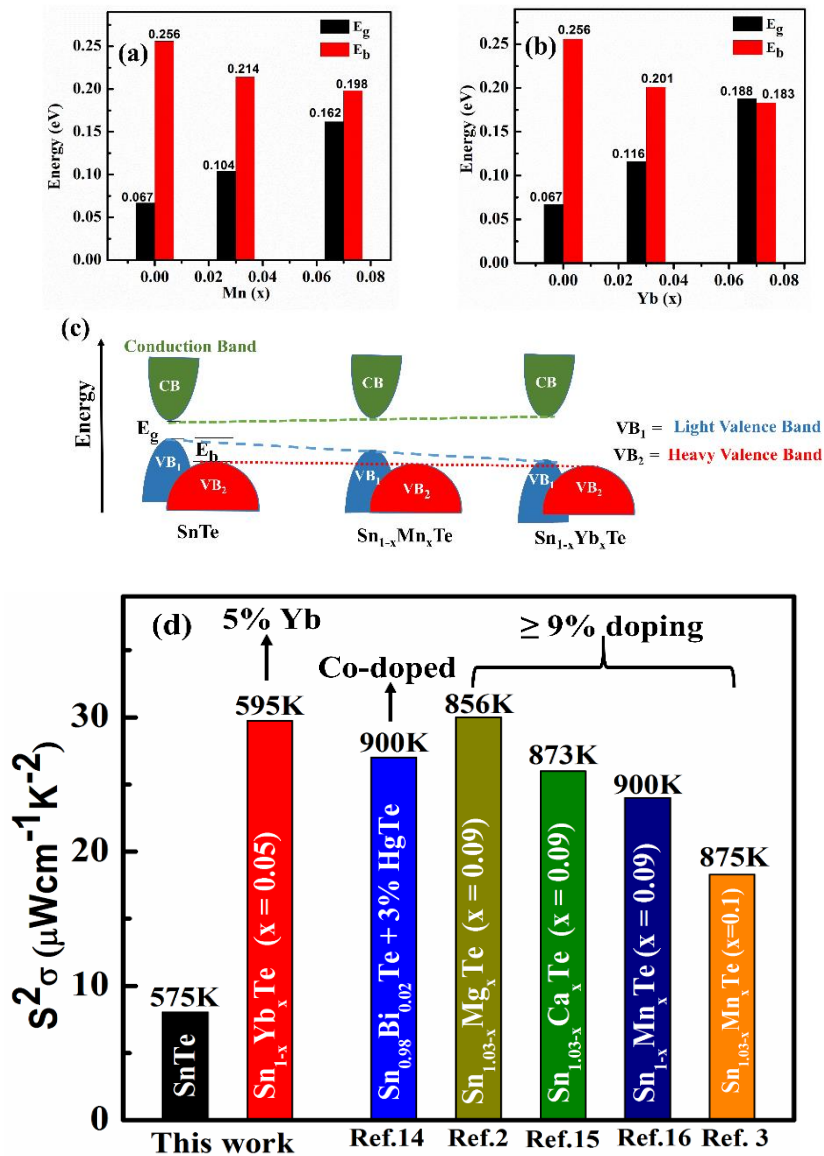

**FIG. 5.** $E_g$ and $E_b$ as a function of (a) Mn and (b) Yb concentration in SnTe; (c) Schematic energy band diagram of SnTe, $Sn_{1-x}Mn_xTe$ and $Sn_{1-x}Yb_xTe$ with conduction band, light and heavy hole valence bands; (d) Comparison of $S^2\sigma$ among pristine SnTe, Yb doped SnTe, $Sn_{0.98}Bi_{0.02}Te+3\%HgTe$, $Sn_{0.94}Mg_{0.09}Te$, $Sn_{0.94}Ca_{0.09}Te$, $Sn_{0.91}Mn_{0.09}Te$ and $Sn_{0.93}Mn_{0.1}Te$.

In summary, doping of ~5% Yb demonstrates an effective band-convergence in SnTe in comparison to doping with Mn and resulting in enhancement of DOS near $E_F$. Our experimental data and first principle calculations support and justify the three-time enhancement of $S^2\sigma$ with Yb doping. Here, Yb acts as an efficient dopant in SnTe due to heavy atomic mass and the resulting SOC. We emphasize that even a small amount (~ 5%) of Yb doping can endow a significant performance to SnTe via effective valence band-convergence and improved electronic density of states.

**Acknowledgements**

AS gratefully acknowledges Board of Research in Nuclear Sciences, India for young scientist research award (37(3)/14/02/2015/BRNS) and Indian Institute of Technology Mandi for research facilities. DD acknowledges Department of Science and Technology (DST), India for INSPIRE research fellowship.